\documentclass[twocolumn,a4paper,10pt,reprint,longbibliography,aip]{revtex4-1}

\usepackage{hyperref}

\usepackage{graphicx}
\usepackage{subfigure}
\usepackage{umoline}
\usepackage{subfigure}
\usepackage{latexsym}   
\usepackage{amsmath,amssymb,amsthm}
\usepackage{bm}

\usepackage{color}

\makeatletter
\renewcommand{\p@subsection}{}
\renewcommand{\p@subsubsection}{}
\makeatother

\begin{document}

\title{Microrheology of semiflexible filament solutions based on relaxation simulations}

\author{L. K. R. Duarte}

\affiliation{Departamento~de~F\'isica,~Universidade~Federal~de~Vi\c{c}osa~(UFV),~36.570-900,~Vi\c{c}osa,~MG,~Brazil.}

\affiliation{Instituto Federal de Educa\c{c}\~ao,~Ci\^encia e Tecnologia de Minas Gerais,~35.588-000,~Arcos, MG, Brazil.}

\author{A. V. N. C. Teixeira}

\affiliation{Departamento~de~F\'isica,~Universidade~Federal~de~Vi\c{c}osa~(UFV),~36.570-900,~Vi\c{c}osa,~MG,~Brazil.}

\author{L. G. Rizzi}

\affiliation{Departamento~de~F\'isica,~Universidade~Federal~de~Vi\c{c}osa~(UFV),~36.570-900,~Vi\c{c}osa,~MG,~Brazil.}



\begin{abstract}
We present an efficient computational methodology to obtain the viscoelastic response of dilute solutions of semiflexible filaments. By considering an approach based on the fluctuation-dissipation theorem, we were able to evaluate the dynamical properties of probe particles immersed in solutions of semiflexible filaments from relaxation simulations with a relatively low computational cost and higher precision in comparison to those based on stochastic dynamics. We used a microrheological approach to obtain the complex shear modulus and the complex viscosity of the solution through its compliance which was obtained directly from the dynamical properties of a probe particle attached to an effective medium described by a mesoscopic model, {\it i.e.}, an effective filament model (EFM). The relaxation simulations were applied to assess the effects of the bending energy on the viscoelasticity of semiflexible filament solutions and our methodology was validated by comparing the numerical results to  experimental data on DNA and collagen solutions.

\vspace{0.5cm}

~
\end{abstract}


\maketitle


\section{Introduction}

	Despite of its importance to the well-functioning of almost all
biological specimens~\cite{rizzi2018eac}, 
the assessment of the viscoelastic response of complex solutions of 
semiflexible filaments~\cite{pritchard2014softmatter}, 
{\it e.g.}, collagen, actin, rodlike viruses, amyloid fibrils, microtubules, and DNA, 
is still a difficult task to our current theoretical and computational approaches~\cite{clasen2019softmatter}.
	Contrary to solutions of cross-linked filaments, where a shear protocol can be used to extract the mechanical properties of the networks~\cite{broedersz2014rmp,rizzi2015prl,rizzi2016sm,terentjev2017polymer},
the study of the viscoelastic response of solutions of diluted unentangled filaments relies mainly on indirectly monitoring the stochastic, {\it i.e.}, the fluctuating, dynamics of structures in the system~\cite{cruz2012archcomput,larson2005jrheol}.
	Alternatively, one could consider relaxation approaches which are based on the fluctuation-dissipation theorem (FDT), just as done experimentally in the microrheological characterization of complex solutions~\cite{gittes1997prl,tassieri2010jrheol,head2014pre}.~However, 
	even though computational simulations based on the FDT have been used to obtain the dynamics of systems with ideal networks~\cite{licinio1997pre,licinio1998phylmag,teixeira1999epl}, such relaxation simulations have not been applied to obtain the viscoelastic properties of solutions; that is the focus of the present study.

	Previous computational efforts indicate that the characterization of the viscoelastic response of dilute solutions by first-principles, {\it i.e.}, molecular-based, approaches can be very challenging~\cite{gartner2019macromol}.~Although 
	simulations considering atomistic models might be able to account for detailed molecular interactions including, {\it e.g.}, polymer-solvent interactions and also hydrodynamic effects between different polymeric chains, they can be hardly used to retrieve the relaxation behaviour of the structures in the system that is required to describe the viscoelastic response of the solutions at the experimentally relevant ({\it i.e.},~mesoscopic) time and length scales, {\it e.g.}, miliseconds  and micrometers.
	Even so, such detailed simulations might provide the basis for systematic coarsening procedures~\cite{gartner2019macromol}, but 
 here we restrict ourselves to a 
less ambitious yet
complementary modelling approach that is based on the microrhelogical characterization of the solutions, where the relaxation behaviour of the system is extracted from the dynamics of probe particles immersed in the medium~\cite{waigh2016review}.

	For simplicity, we consider
that the effects of the coupling between micron-sized probe particles and the dilute solution of filaments can be described by the dynamics of tagged beads in the middle of an effective  
mesoscopic model.~Our 
	idea is to assume the simplest constitutive approach 
which lead to relaxation behaviours that are
similar to the ones that are observed for polymeric solutions.~Thus, we focus our attention to
demonstrate how the relaxation
simulations can be used to efficiently extract the non-markovian dynamics of the probe particles that ultimately allows one to determine the frequency-dependent shear moduli and complex viscosity that are characteristic of viscoelastic solutions of semiflexible filaments.

	The reminder of the paper is as follows.~First, 
in Sec.~\ref{viscoelasticity_sol}, we review some of the relevant theoretical and computational aspects related to
the rheology and microrheology of solutions of unentangled filaments.~In
	Sec.~\ref{methods} we describe a simple mesoscopic constitutive model that is used to mimic the effective
response of the solutions and present the  numerical methods commonly adopted to perform 
stochastic simulations.
	In Sec.~\ref{relax_approach} we discuss the approach based on relaxation simulations,
which allow us to extract the viscoelastic properties of solutions from the dynamics of the 
probe particles described by the effective mesoscopic model.
	Also, we validate our methodology by including
a comparison between the results obtained from the relaxation simulations and the stochastic simulations, as well as a comparison between the numerically obtained results and the experimental data obtained for solutions of
polyelectrolytes.~The
	results for semiflexible filaments are presented in Sec.~\ref{semiflexible},
where we investigate the effects of bending energies on the viscoelasticity of dilute solutions
and we also include comparisons between the numerical results and the experimental data obtained 
for DNA and collagen macromolecules.

\section{Viscoelasticity of unentangled solutions}
\label{viscoelasticity_sol}

\subsection{Viscoelastic response functions}

	Experimentally, one can characterize the mechanical properties of viscoelastic fluids
by considering, {\it e.g.},~steady-state shearing experiments~\cite{larsonbook}, where the time-dependent stress $\sigma(t)$ of the viscoelastic material is related to its relaxation modulus $G(\tau)$ as~\cite{ferrybook,rubinstein}
\begin{equation}
\sigma(t) = \int_{-\infty}^{t} dt' \, G(t-t')\, \dot{\gamma}(t') ~~,
\label{sigmat}
\end{equation}
with $\dot{\gamma}$ being the shear rate.~Also, 
	one can consider small-amplitude oscillatory shearing experiments~\cite{larsonbook}
with $\gamma(t)$ being an oscillatory function so that
the viscoelastic response of the fluid is given by the complex shear modulus, $G^{*}(\omega)=G^{\prime}(\omega) + i G^{\prime \prime}(\omega)$, where $G^{\prime}(\omega)$ and $G^{\prime \prime}(\omega)$ correspond to the storage and the loss modulus, respectively.~At
	the linear viscoelastic (LVE) response regime, both experimental techniques should provide the same information, 
as the complex modulus is directly related to $G(\tau)$ via a Fourier transform~\cite{larsonbook,doisoftmatter}, that is,
\begin{equation}
G^{*}(\omega) = i \omega \int_{0}^{\infty} d\tau' \, G(\tau') \, e^{-i\omega \tau'} ~~.
\label{Gstar_Gtau}
\end{equation}
	For viscoelastic solutions, it is also convenient to obtain the complex viscosity~\cite{larsonbook}, $\eta^{*}(\omega)=\eta^{\prime}(\omega) - i \eta^{\prime \prime}(\omega)$, which is related to the complex modulus as $G^{*}(\omega) = i\omega \, \eta^{*}(\omega)$, so that $\eta^{\prime}(\omega) = G^{\prime \prime}(\omega)/\omega$ and $\eta^{\prime \prime}=G^{\prime}(\omega)/\omega$.
	At low frequencies, the loss modulus of viscoelastic solutions is expected to be proportional to the frequency, {\it i.e.},~ $G^{\prime \prime}(\omega) \propto \omega$, so the viscosity $\eta^{\prime}(\omega)$ should be independent of the frequency and is related to the relaxation modulus as~\cite{doisoftmatter} 
\begin{equation}
\eta_0^{\,} = \lim_{\omega \rightarrow 0}\eta^{\prime}(\omega) = \int_{0}^{\infty} d\tau' G(\tau') ~~,
\label{eta0}
\end{equation}
which is, according to the Cox-Merz rule, equivalent to the steady-state viscosity $\eta(\dot{\gamma}) = \sigma(\dot{\gamma})/\dot{\gamma}$ that is obtained for low shear rates at the LVE regime~\cite{li2005jdisperscitech}.

\subsection{Microrheology}

	Alternatively, the viscoelasticity of complex solutions can be obtained through microrheological techniques~\cite{squires2010annrev,waigh2016review}, which are based on relationships between the viscoelastic response of the material and the dynamics of probe particles immersed in it (see,~{\it e.g.},~Ref.~\cite{tassieri2015microrheology}).~In particular, 
	one can explore passive experimental approaches ({\it e.g.},~particle tracking videomicroscopy or dynamic light scattering; see Ref.~\cite{waigh2016review} for a review) to extract the mean-squared displacement (MSD) $\langle \Delta r^2 (\tau) \rangle_{a}^{\,}$ of probe particles with radius $a$ and relate it to the compliance $J(\tau)$ of the solution through a generalized Stokes-Einstein 
relationship~\cite{squires2010annrev,wirtz1998rheolacta,mason2000rheolacta}, that is,
\begin{equation}
J(\tau) = \frac{3 \pi a}{d k_B T} \langle \Delta r^2 (\tau) \rangle_{a}^{\,}
 ~~,
\label{compliance}
\end{equation}
where $d$ is the euclidean dimension of the random walk, $k_B$ is the Boltzmann's constant, and $T$ is the absolute temperature of the medium.
	A simple way to understand this relationship is by considering the diffusion of the probe particles at later times, i.e, at times $\tau$ that are longer than the longest relaxation time $\tau_f$ of the solution.
	In that case one should observe a normal diffusive behaviour where the MSD is given by $\langle \Delta r^2 (\tau) \rangle_{a}^{\,}  = 2 d D_{a} \tau$ with 
\begin{equation}
D_{a} = \frac{k_BT}{6 \pi a \eta_0^{\,} }~~,
\label{diff_coef_a}
\end{equation}
so that the right hand side of Eq.~(\ref{compliance}) yields $\tau/\eta_0^{\,}$, which is the compliance $J(\tau)$ that one would measure from creep-compliance experiments, {\it i.e.}, which is independent of the radius $a$ of the probe particles.~Importantly, the above expression is valid only for relatively large and isolated particles which effectively probe the viscosity $\eta_0^{\,}$ of the 
	solution~\cite{squires2010annrev}
	(usually, micron-sized beads are chosen to probe the LVE response of polymeric solutions~\cite{rizzi2018eac}).

	At the LVE regime, microrheology and rheology are expected to give the same information about the viscoelastic behaviour of the solution.~Also, one should note that the relaxation modulus is linked to the compliance $J(\tau)$ of the solution through a convolution~\cite{ferrybook},
\begin{equation}
\int_{0}^{\tau} G(\tau-\tau')J(\tau') d\tau' = \tau ~~,
\label{relation_GtauJtau}
\end{equation}
and one can evaluate the complex shear modulus directly from the Fourier transform of the compliance $\hat{J}(\omega)$ as
\begin{equation}
G^*(\omega)=\dfrac{1}{i\omega \hat{J}(\omega)} ~~.
\label{complexG_J}
\end{equation}

\subsection{Stochastic dynamics and relaxation of polymers}

	Theoretically, the viscoelastic response of diluted unentangled filament solutions can be evaluated through estimates of the intrinsic relaxation modulus~\cite{doiedwards} $[G(\tau)]$.
	Based on polymer physics, one may resort to a heuristic argument that the relaxation times will depend on the relaxation of partial segments of the filaments so that an approximated expression for the relaxation modulus can be written as~\cite{rubinstein,doisoftmatter}
\begin{equation}
G(\tau) \propto n_f k_B T \left( \frac{\tau}{\tau_f} \right)^{-\alpha} e^{-(\tau/\tau_f)}~~,
\label{relaxation_modulus_f}
\end{equation}
where $n_{f}$ is the number density of filaments, $\tau_{f}$ is the longest relaxation time of the filaments in solution,
and $\alpha$ is an exponent that characterizes the effective flexibility of the filaments~\cite{rubinstein,doisoftmatter}.
	By inserting $G(\tau)$ into expression~(\ref{Gstar_Gtau}) one finds that the frequency-dependent shear moduli present a power-law behaviour with the same exponent, that is, $G^{\prime}(\omega) \propto  G^{\prime \prime}(\omega) \propto \omega^{\alpha}$, at intermediate frequencies ({\it i.e.}, $\omega > \omega_f$ with $\omega_f=1/\tau_f$).

	Interestingly, by assuming the same heuristic principle, numerical results obtained from 
simulations using single chains~\cite{binder1991jcp,pasquali2001pre}
have also suggested that the relaxation behaviour of dilute solutions at intermediate times, {\it i.e.},~$\tau < \tau_f$, should be somewhat related to the stochastic dynamics of the monomers in the middle of the polymeric chains.~It 
	seems that the intrinsic relaxation modulus should show a power-law behaviour which display the same characteristic exponent of the subdiffusive behaviour observed for the MSD of the monomeric units in the polymeric chain, that is,~$\langle \Delta r^2 (\tau) \rangle_{m}^{\,} \propto \tau^{\alpha}$, with $\alpha<1$.
	For semiflexible chains, in particular, recent molecular dynamics simulations~\cite{binder2016jcp} indicate that the MSD of monomeric units display an exponent $\alpha = 3/4$, which agreed with several theoretical approaches~\cite{rubinstein,shankar2002jrheol} and experimental evidence~\cite{clasen2019softmatter,tassieri2012newj,sarmiento2012EPJE,krajina2017acs} in the literature.~Numerical 
simulations presented in Ref.~\cite{binder2016jcp}
	confirmed the results obtained in Refs.~\cite{pasquali2001pre,dimitrakopoulos2001pre,spakowitz2014pre} that both bond autocorrelation and end-to-end vector correlation functions (which are extracted from an isolated chain and can be related to the intrinsic relaxation modulus), display a power-law behaviour that lead to a similar exponent.

\begin{figure}[!b]
	\centering
	\includegraphics[scale=0.46]{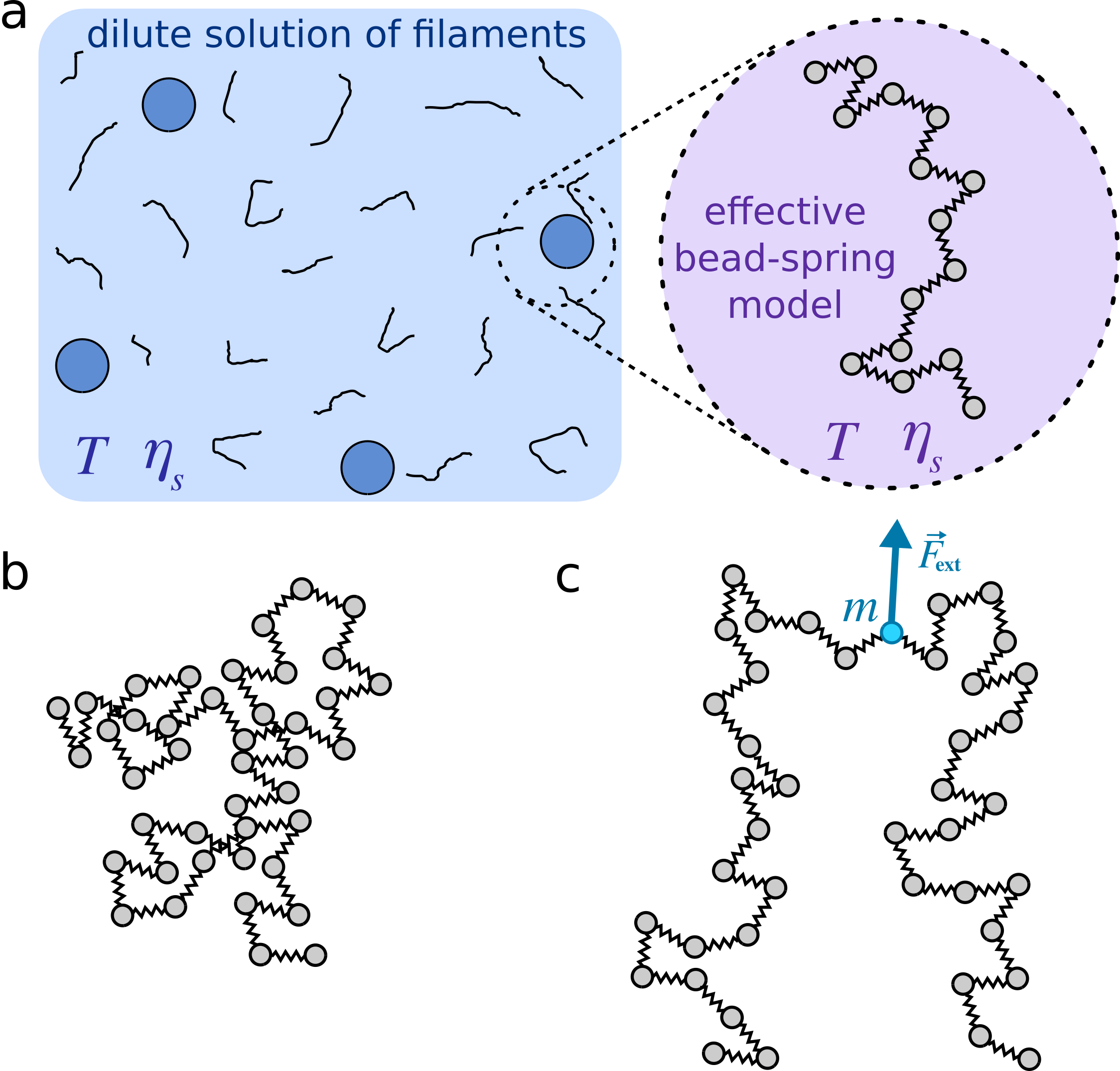}
	\caption{(a)~The dynamics of a probe particle (blue circle) immersed in a mesoscopic region of the solution of diluted filaments (left panel) is effectively described by the dynamics of a bead (grey circles) in a bead-spring model (right panel).
	The effective filament model (EFM) at the right panel is used as a constitutive model to describe
the relaxation of solutions of both flexible and semiflexible filaments.
	The non-markovian dynamics of beads in the EFM depends on their interactions, which are determined by effective elastic ($\kappa$) and bending ($\kappa_b$) constants defined in Eqs.~(\ref{harmonic}) and~(\ref{bending}), respectively.
	The link between the EFM (right panel) and the viscoelastic properties of the corresponding solution (left panel) is made through 
the diffusion coefficients $D_a$, Eq.~(\ref{diff_coef_a}), and $D_f$, Eq.~(\ref{longtime_diff_coeff}), which are defined in terms of the radius $a$ of the probe particle.
	(b) Typical configuration of the EFM in the overdamped Langevin 
approach, where the dynamics of its beads is stochastic.
 (c)~Configuration in the relaxation approach, where an external force $\vec{F}_{\text{ext}}$ is used to pull the $m$-th bead 
placed in the middle of the EFM.}
	\label{fig1}
\end{figure}

	As discussed in Refs.~\cite{panja2010jstatmech_l02001,panja2010jstatmech_p06011},
one can actually use a generalized theoretical Langevin approach to recover the same relaxation behaviour given by Eq.~(\ref{relaxation_modulus_f}) with $\alpha=1/2$ ({\it i.e.},~for Rouse chains) in the overdamped regime by considering the non-markovian dynamics of a tagged particle in the middle of a flexible chain.
	In addition, Reference~\cite{rizzi2020jrheol} shows that it is possible to consider a similar generalized 
Langevin approach to link the shear moduli of a viscoelastic material to the dynamics of a probe particle through a 
microrheological approach based on Eq.~(\ref{compliance}).
	Hence, in order to demonstrate the usefulness of the relaxation simulations introduced here without having to resort to atomistic simulations, we consider 
a mesoscopic constitutive model that effectively describes the non-markovian dynamics of a probe particle immersed in a mesoscopic region of a dilute solution of filaments, just as shown in Fig.~\ref{fig1}(a).


\section{Numerical methods}
\label{methods}

Before getting into how one can obtain the viscoelastic response functions from the relaxation simulations, we introduce in this Section the aforementioned mesoscopic model used to describe the coupling between the probe particle and the dilute solution of filaments, as well as the numerical procedures used to obtain its dynamical properties from stochastic simulations which are explored latter to validate our relaxation-based approach.

\subsection{Effective filament model (EFM)}
\label{filamentmodel}

	As illustrated in Fig.~\ref{fig1}(a), 
we represent a probe particle in a mesoscopic region of the solution of filaments by 
an effective filamentous structure ({\it i.e.}, the EFM) which is modelled 
by a simple bead-spring model with $N$ beads in an implicit solvent that has the same
properties of the original solution, {\it i.e.}, its viscosity
$\eta_s$ and temperature $T$.
	The beads in such EFM are coupled to their nearest neighbours by an interaction potential 
that can include contributions from both harmonic ($U_{\text{h}}$) and bending ($U_{\text{b}}$) energies.

	The harmonic interaction potential of the whole effective filament is written as
\begin{equation}
U_{\text{h}} = \frac{\kappa}{2} \sum_{j=1}^{N-1} \left(\vec{r}_{j+1}-\vec{r}_{j}\right)^2~~,
\label{harmonic}
\end{equation}
where $\vec{r}_{j}$ is the position vector of the $j$-th bead
and $\kappa$ is the effective elastic constant.~One
	can relate the pre-factor in Eq.~(\ref{harmonic}) to the pre-factor of the Gaussian chain model~\cite{doiedwards}, so that $\kappa = 3 k_B T / b^2$, where 
$b$ is a parameter that sets the length scale ({\it e.g.},~nm) and the strength of the harmonic interaction.

	For the bending interaction potential we assume its discretized approximation (see, {\it e.g.},~Ref.~\cite{likhtman2007macromol}), which is evaluated as the sum of local curvatures along the EFM and is given by
\begin{equation}
U_{\text{b}} = \frac{\kappa_b}{2} \sum_{j=2}^{N-1} \left(\vec{r}_{j-1}-2\vec{r}_{j}+\vec{r}_{j+1}\right)^2~~,
\label{bending}
\end{equation}
where $\kappa_b = E/b^4$ is the bending constant,
with $E$ being a parameter that sets the bending stiffness.
	Both constants $\kappa$ and $\kappa_b$ are given in units of force per length, {\it e.g.},~pN/nm.
	The value of $\kappa_b$ can be approximately related to the persistent length of the filament 
$L_p$, since it should be proportional to $A \equiv E^{\prime}/ b$, with $E^{\prime}=E/k_BT$
({\it e.g.}, if $E$ is given in pN.nm$^3$ and $b$ in nm, $E^{\prime}$ is given in nm$^2$, and $L_p$ and $A$ in nm).

\subsection{Stochastic simulations}
\label{stochastic_sim_sec}

	First, in order to validate 
 the relaxation simulations, we compare it to stochastic, {\it i.e.}, Brownian dynamics, simulations,
which consist in solving
numerically the overdamped Langevin equation.~For
	 the $i$-th bead in the EFM, such equation is written as
\begin{equation}
\frac{\partial \vec{r}_i}{\partial t} = \frac{1}{\zeta} \left( 
\vec{F}_i
+ \vec{f}_{a} \right) ~~,
\label{overdampedBD}
\end{equation}
where $\vec{f}_{a}$ is a random force due to interaction of the bead with the implicit effective solvent,
$\zeta$ is a 
time-independent friction coefficient, 
and 
$\vec{F}_i$ 
is the total force exerted on the $i$-th bead which is determined from the interaction potentials defined by~Eqs.~(\ref{harmonic}) and~(\ref{bending}),
{\it i.e.},~$\vec{F}_i=- \nabla_i ( U_{\text{h}} + U_{\text{b}} )$, 
with $\nabla_i = \partial_{x_i} \hat{x} + \partial_{y_i} \hat{y} + \partial_{z_i} \hat{z}$.

	In practice, one have $3N$ coupled differential equations defined as in Eq.~(\ref{overdampedBD}), which are discretized and solved numerically by considering the Euler integration scheme, so that the position vector of the $i$-th bead at a time $t+\Delta t$ is given by
\begin{equation}
\vec{r}_i(t+\Delta t) = \vec{r}_i(t) +  \frac{\Delta t}{\zeta} 
\left(
\vec{F}_{i}
+ \vec{f}_{a} \right)
~~,
\label{langevin_integrated_brownian}
\end{equation}
where the $k$-th component of the random force is evaluated as~\cite{gillespie1993amjphys}
\begin{equation}
f_{a,k} = \sqrt{ \frac{2 \zeta k_B T}{\Delta t} } \, \text{N}_{k}(0,1) ~~,
\label{randomforce}
\end{equation}
with $\text{N}_{k}(0,1)$ (for $k=x$, $y$, or $z$) 
being independent 
random variables obtained from a gaussian distribution with zero mean and variance equal to one.
	We assume that the value of the effective friction coefficient $\zeta$ depends on the radius $a$ of the probe particle and is determined by the Stokes-Einstein relation, that is,
\begin{equation}
\zeta = \frac{k_B T}{D_0} = 6 \pi a \eta_s~~,
\label{selfdifcoef}
\end{equation}
where $D_0=k_BT/6 \pi a \eta_s$ defines the diffusion coefficient of a non-connected bead, {\it i.e.}, a probe particle with radius $a$ freely diffusing
in a solvent with viscosity $\eta_s$.

	Hence, we impose that the dynamics of probe particles in the solution of filaments given by $\langle \Delta r^2(\tau) \rangle_a$ can be effectively characterized by the fluctuations in the position of the beads of the EFM, 
which are quantified by their mean-squared displacement,
\begin{equation}
\langle \Delta r^2(\tau) \rangle = \langle \left[ \vec{r}(\tau+t_0) - \vec{r}(t_0) \right]^2 \rangle ~~,
\label{deltaR2}
\end{equation}
where $\langle \dots \rangle$ denote averages over both 
$N_T$ beads and $M$ realizations of the numerical experiment.
	The initial configuration in each numerical simulation corresponds
to a fully stretched chain with the beads separated by a distance $b$,
and the averages are evaluated only after a thermalization period of time 
$t_0$.

	The time-dependent diffusion coefficient $D(\tau)$ of the beads in the EFM can be 
retrieved from the time derivative of the MSD of the bead's position, that is
\begin{equation}
D(\tau) = \frac{1}{2 d} \frac{\partial \langle \Delta r^2(\tau) \rangle}{\partial \tau} ~~.
	\label{diffconstdef}
\end{equation}
	For all simulations we consider that the euclidean dimension is $d=3$.
	It is worth mentioning that, in order to avoid boundary effects on $\langle \Delta r^2(\tau) \rangle$ and $D(\tau)$, 
we consider that the average value is 
evaluated over the $N_T=N - 2N_E$ beads which are 
centrally localized in the chain, {\it i.e.},~excluding $N_E$ beads on each side
of the EFM.

\begin{figure}[!t]
	\centering
	\includegraphics[scale=0.54]{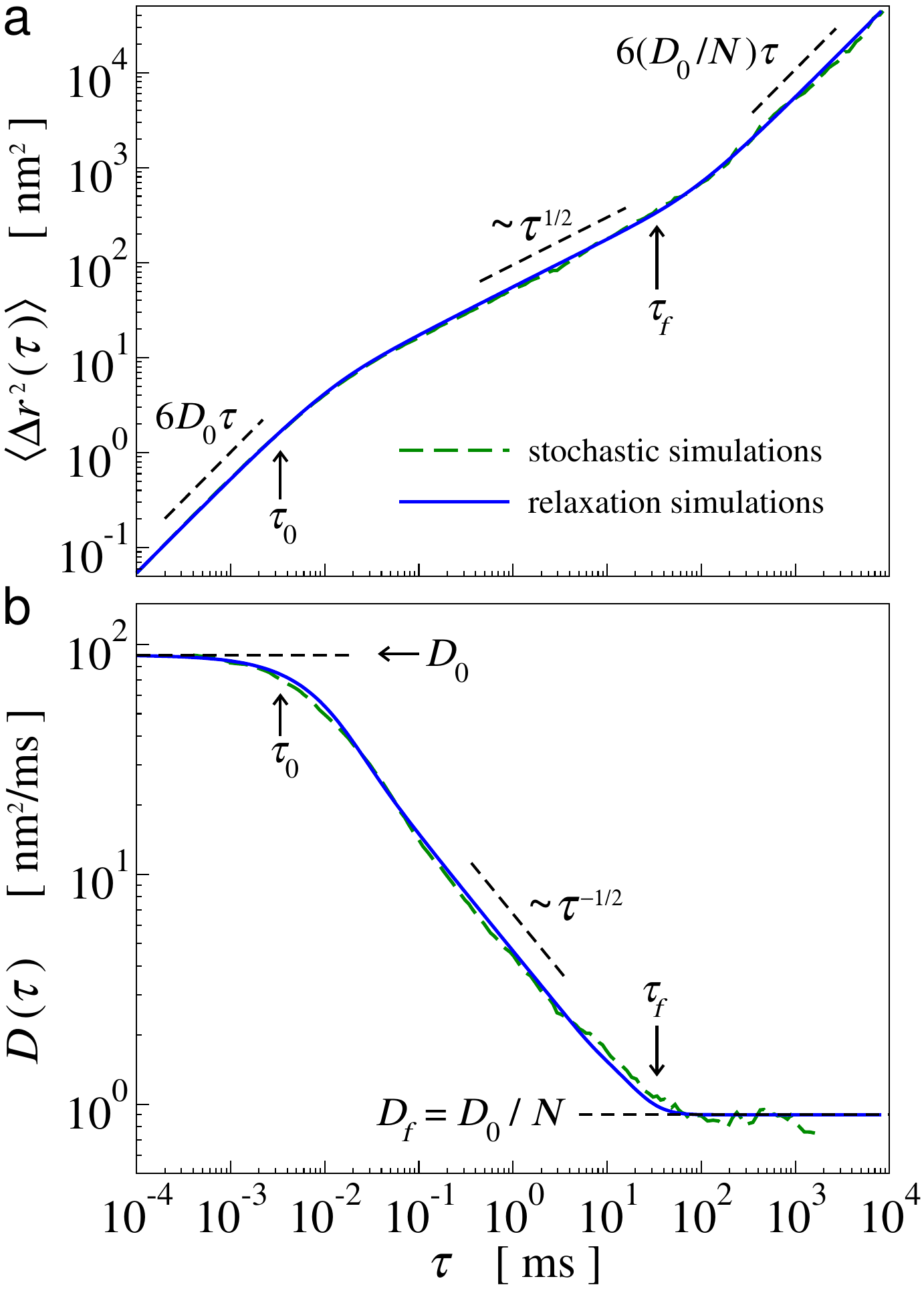}
	\caption{(a) Mean-squared displacement $\langle \Delta r^2(\tau)\rangle$, and (b) time-dependent diffusion coefficient $D(\tau)$, as function of time $\tau$,
for a flexible EFM ({\it i.e.}, no bending energy, $\kappa_b=0\,$pN/nm).
	Long-dashed green lines correspond to stochastic simulations,
 while straight blue lines correspond to relaxation simulations.
	Short-dashed black lines indicate the two normal diffusive behaviours where $\langle \Delta r^2(\tau) \rangle \propto \tau$ and $D(\tau)$ is constant, and the intermediate subdiffusive behaviour with $\langle \Delta r^2(\tau)\rangle \propto \tau^{1/2}$ and $D(\tau) \propto \tau^{-1/2}$, as expected from Rouse dynamics~\cite{doiedwards} (see text for details).
	For both numerical approaches the results were obtained 
with the arbitrary parameters
$N=100$, 
$D_0=90\,$nm$^2$/ms,
$\kappa = 1.38\,$pN/nm ($b=3\,$nm), 
and
$T=300\,$
($k_BT=4.142\,$pN.nm),
so that
$\tau_0 = k_B T /(\pi^2 \kappa D_0) 
\approx 0.0034\,$ms,
and $\tau_f = \tau_0 N^2 \approx 34\,$ms.
	For the stochastic simulations, the results correspond to average values obtained from $M=100$ realizations with $N_E=7$ after a thermalization of 
$t_0=5 \times 10^2\,$ms ({\it i.e.},~ 5 $\times 10^6$ steps with $\Delta t=10^{-4}\,$ms).
	For the relaxation simulations a constant external force $F_0=1\,$pN and the same $\Delta t$
were considered but no thermalization was required.}
	\label{fig2}
\end{figure}

\subsection{Flexible EFM}
\label{flexibleEFM}

	In order to illustrate the dynamics of the EFM and validate our relaxation approach
we present in Fig.~\ref{fig2}  a comparison between the two methods for
the MSD and the time-dependent diffusion coefficient obtained for 
a flexible EFM, {\it i.e.},~without bending energy (the relaxation approach will be described in the next section).
Without bending energies, the dynamics of the beads in the EFM 
can be quantitatively described by the Rouse model~\cite{rubinstein,doiedwards} and, as expected, the MSD displays two normal diffusion regimes: one for times shorter than 
$\tau_0 = k_B T /(\pi^2 \kappa D_0)$,
with $\langle \Delta r^2 (\tau) \rangle = 6 D_0 \tau$, 
which corresponds to the free-like displacements of the beads;
and the other 
for times longer than $\tau_{f} = \tau_0 N^2$, with 
$\langle \Delta r^2 (\tau) \rangle = 6 ( D_0 / N) \tau$,
which corresponds to the diffusion of the centre of mass of the EFM.
	Also, the Rouse model predicts an intermediate regime with a characteristic subdiffusive
anomalous behaviour~\cite{doiedwards}, where 
$\langle \Delta r^2(\tau) \rangle = \sqrt{36 k_BT D_0 /(\pi \kappa )} \, \tau^{1/2}$.~As
	shown in Fig.~\ref{fig2}(b), those regimes are better identified by
the time-dependent diffusion coefficient $D(\tau)$, which shows a transient power-law 
regime,
 {\it i.e.},~$D(\tau) =  \sqrt{k_B T D_0 /(4 \pi \kappa)} \, \tau^{-1/2}$, 
between two plateaus,
one with $D(\tau) = D_0$ at times shorter than 
$\tau_0\approx 0.0034\,$ms,
 and the other with
$D(\tau) = D_f = D_0/N$, at times longer than 
$\tau_f\approx 34\,$ms.


\section{Relaxation approach}
\label{relax_approach}

	In this Section we describe the theory and the numerical procedures involved in the relaxation simulations that are used to obtain the dynamical properties of the EFM, as well as how those properties can be used to provide numerical estimates for the shear moduli and the complex viscosity of the corresponding diluted filament solutions.

\subsection{Relaxation simulations based on the FDT}
\label{relax_sim}

	In the relaxation approach, the MSD $\langle \Delta r^{2}(\tau) \rangle$ and the time-dependent diffusion coefficient $D(\tau)$ of the beads in the EFM are evaluated from a relation that comes from the fluctuation-dissipation theorem (FDT)~\cite{doisoftmatter,doiedwards}.

	Importantly, the use of the relaxation simulations based on FDT is restricted to the linear response regime, which means that the intensity of the external force is relatively weak but large 
enough so that one can neglect the random thermal forces $\vec{f}_a$.
	In this case one can solve the $3N$ coupled differential equations by using a Euler integration scheme similar to Eq.~(\ref{langevin_integrated_brownian}), but assuming that $\vec{f}_a$ are close to zero, so that
\begin{equation}
\vec{r}_i(t+\Delta t) = \vec{r}_i(t) +  \frac{\Delta t}{\zeta} 
\left(
\vec{F}_{i}
+ \delta_{im} \vec{F}_{\text{ext}} \right) ~~,
\label{langevin_integrated}
\end{equation}
where the Kronecker's delta $\delta_{im}$ indicates that the constant external force 
$\vec{F}_{\text{ext}} = F_0\,\hat{z}$ is applied only to the $m$-th bead in the middle of the EFM, 
as illustrated in Fig.~\ref{fig1}(c).

	In practice, the FDT can be used to link the displacement $\Delta z(\tau)$ of the $m$-th bead driven by an external force to the fluctuations on its position at equilibrium as~\cite{doisoftmatter}
\begin{equation}
\Delta z(\tau) = \left[ z_{m}(\tau)-z_{m}(0) \right] = \chi_{zz}^{\,}(\tau) F_{0} ~~,
\end{equation} 
where $\chi_{zz}^{\,}(\tau)$ is a linear response function given by
\begin{equation}
\chi_{zz}^{\,}(\tau) = \frac{1}{2 k_B T} \langle \Delta z^2(\tau) \rangle ~~.
\end{equation}
	Hence, one can estimate the MSD of the beads in $d$ dimensions as
\begin{equation}
\langle \Delta r^2(\tau)\rangle = \frac{2 d k_BT}{F_0} \left[ z_{m}(\tau)-z_{m}(0) \right]~~~.
\label{deltaR2relax}
\end{equation}
	Also, one can retrieve the time-dependent diffusion coefficient $D(\tau)$ by 
derivating Eq.~(\ref{deltaR2relax}) just as prescribed by Eq.~(\ref{diffconstdef}), which yields
\begin{equation}
D(\tau) = \frac{k_BT}{F_0} \upsilon_{m,z}(\tau) ~~~,
\label{relaxDtau}
\end{equation}
where $\upsilon_{m,z}(\tau)$ is the velocity of the $m$-th bead, which can be directly obtained from the numerical integration scheme.

	As one can see in Fig.~\ref{fig2}, the results obtained from the relaxation approach with Eqs.~(\ref{deltaR2relax}) and~(\ref{relaxDtau}) display a good agreement to those obtained from the stochastic simulations.
	It is worth mentioning that, since one does not have to compute averages over $M$ realizations and it does not require the thermalization step ({\it i.e.},~the initial configuration corresponds to a fully stretched chain placed along a direction that is perpendicular to $z$ with a separation $b$ between beads), the numerical approach based on relaxation dynamics is far more efficient than the one based on stochastic dynamics.
	For instance, the results obtained from relaxation simulations presented in Fig.~\ref{fig2} took less than a minute to be produced, while the simulations using the stochastic approach required several hours.
	Also, the numerical data obtained from relaxation simulations
is not noisy as those obtained from the stochastic simulations.
	That is very convenient since, as we discuss in the following, one have to compute Fourier transforms of $\langle \Delta r^2(\tau) \rangle$ in order to extract the viscoelastic properties of the solutions.

	It is worth noting that, since the FDT expressions are very general, the relaxation approach based on Eqs.~(\ref{deltaR2relax}) and~(\ref{relaxDtau}) could be applied to models other than the EFM defined in Sec.~\ref{filamentmodel}, {\it e.g.}, molecular-based models with explicit solvent, just as it is done in experiments~\cite{tassieri2010jrheol}.
	Even so, just to illustrate the determination of the viscoelastic functions from those equations, we discuss in the next Section how one can explore the relationship between the EFM and the Rouse model to describe the experimental data.


\subsection{Viscoelastic properties}
\label{visco_prop}

	As discussed in Sec.~\ref{viscoelasticity_sol}, the viscoelastic properties of the filament solution are characterized by the complex shear modulus $G^*(\omega)=G^{\prime}(\omega) + i G^{\prime \prime}(\omega)$, which can be evaluated from the Fourier transform of the compliance of the solution $J(\tau)$ based on Eq.~(\ref{complexG_J}).
	The idea of using an approach based on microrheology is that one can obtain 
$J(\tau)$ directly from the dynamics of probe particles, {\it e.g.}, from the MSD of the beads in the EFM, 
through Eq.~(\ref{compliance}).

\begin{figure*}[!ht]
	\centering
	\includegraphics[scale=0.455]{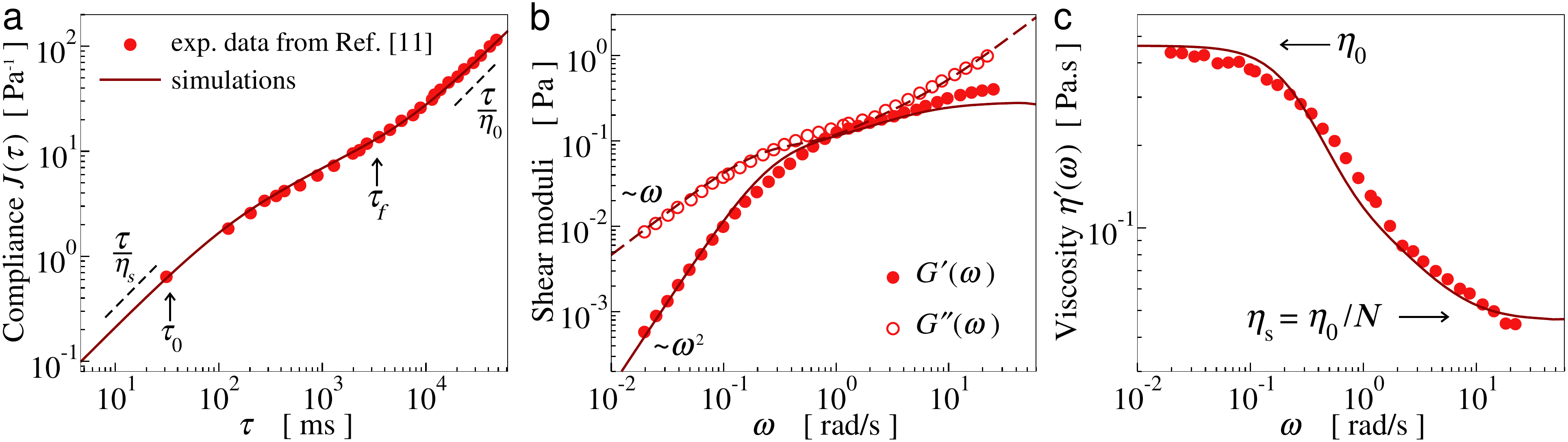}
	\caption{Comparison between the 
experimental data (circles) on flexible polyelectrolyte chains
extracted from Ref.~\cite{tassieri2010jrheol}
and the numerical results obtained from the relaxation simulations (continuous lines).
(a) Compliance $J(\tau)$, (b) storage modulus $G^{\prime}(\omega)$ (filled circles) and loss modulus $G^{\prime \prime}(\omega)$ (open circles), and (c) viscosity, $\eta^{\prime}(\omega)$.
	Numerical estimates for the MSD of the EFM beads $\langle \Delta r^2(\tau) \rangle$ were obtained via Eq.~(\ref{deltaR2relax}) from relaxation simulations implemented with 
$F_0=1\,$pN,
$N  = 10$, 
$D_0 = 1.69  \,$nm$^2$/ms,
$\kappa = 7.28 \times 10^{-3}\,$pN/nm ($b \approx 41\,$nm), 
and $\Delta t   = 0.1\,$ms 
	(the specific values of $D_0$ and $\kappa$ were determined respectively through Eqs.~(\ref{longtime_diff_coeff}) and~(\ref{kappa_of_a}) by setting the radius of the probe particles equal to $a=2.8\,\mu$m as in 
Ref.~\cite{tassieri2010jrheol}).
	The match between the compliance $J(\tau)$ obtained from our numerical simulations through Eq.~(\ref{Jtau_nf}) and the experimental data of Ref.~\cite{tassieri2010jrheol} was done by 
considering 
$\eta_0^{\,}\approx 460\,$mPa.s,
$\eta_s^{\,} \approx 46\,$mPa.s,
$T\approx 25^{\text{o}}$C ($k_BT= 4.114\,$pN.nm),
and $\tau_f\approx 3.38\,$s ({\it i.e.}, $\tau_f \approx 3.38 \times 10^3\,$ms). Those values
yield a number density equal to $n_f=2.5 \times 10^{13}\,$cm$^{-3}$ that is close to the value 
 $n_{\text{PAM}}^{\,}$ obtained from the nominal concentration of 0.07\% w/w used in the experiments (see text for details).
	The complex shear modulus, $G^{*}(\omega)$, and the viscosity, $\eta^{\prime}(\omega)=G^{\prime \prime}(\omega)/\omega$, were obtained from $J(\tau)$ via Eq.~(\ref{complexG_J}) using the numerical method described in Ref.~\cite{evans2009pre}.
}
\label{fig3}
\end{figure*}

	For the EFM, in particular, the first task is to obtain its effective parameters ({\it i.e.}, $N$, $D_0$, $\kappa$, and $\kappa_b$)
in order to describe the full behaviour of all the viscoelastic functions, {\it i.e.}, $J(\tau)$, $G^{\prime}(\omega)$, $G^{\prime \prime}(\omega)$, $\eta^{\prime}(\omega)$, by considering only a minimal experimentally available information, {\it e.g.}, $T$, $\eta_s$, $\eta_0$, and $\tau_f$.
	In order to illustrate how that can be done, we first consider dilute solutions of polyelectrolyte chains.
	In particular, in Fig.~\ref{fig3} we include a comparison between the results obtained from our relaxation simulations and the experimental data presented in Ref.~\cite{tassieri2010jrheol} on a solution of polyacrylamide (PAM) chains.

	As discussed in Sec.~\ref{flexibleEFM} and illustrated by the numerical results presented in Fig.~\ref{fig2}, the dynamics of the beads in the flexible EFM can be well described by the Rouse model, so that the limiting values for the time-dependent diffusion coefficient $D(\tau)$ are given by $D_0$ for $\tau \ll \tau_0$ (see Eq~(\ref{selfdifcoef})), and by
\begin{equation}
D_f = \frac{D_0}{N} = \frac{k_BT}{6 \pi a \eta_s N} ~~,
\label{longtime_diff_coeff}
\end{equation}
for $\tau \gg \tau_f$, where the longest relaxation time of the EFM is given by the Rouse relaxation time~\cite{doiedwards}, that is,
\begin{equation}
\tau_f = \frac{ k_B T }{\pi^2 \kappa D_0  } N^2 ~~.
\label{tauf_Rouse}
\end{equation}
	Hence, by assuming that the diffusion coefficient $D_f$ given by Eq.~(\ref{longtime_diff_coeff}) should be equal to the diffusion coefficient $D_a$ of a probe particle with radius $a$ defined by Eq.~(\ref{diff_coef_a}), one can readily identify that 
the relative viscosity of the solution, $\eta_r^{\,}$, is related to the effective number of beads $N$ of the EFM as
\begin{equation}
\eta_r^{\,} = \frac{\eta_0^{\,}}{\eta_s^{\,}} = N~~. 
\label{relative_visco}
\end{equation}
	This expression is very useful since it allows one to estimate the effective number of beads $N$ which the EFM needs
in order to correctly describe the experimental data.~For instance, 
	the value $N = 10$ can be inferred from Fig.~\ref{fig3}(c)
by realizing that the frequency-dependent viscosity $\eta^{\prime}(\omega)$ is given approximately by 
$\eta_{\infty} = \lim_{\omega \rightarrow \infty} \eta^{\prime}(\omega) = \eta_0^{\,} / N = \eta_s^{\,}$ at high frequencies, and by $\eta_0^{\,} = \lim_{\omega \rightarrow 0} \eta^{\prime}(\omega)$, 
at low frequencies.~Alternatively, 
	one can also obtain the value of $N$ from Eq.~(\ref{longtime_diff_coeff}) by measuring the diffusion coefficients $D_0$ and $D_f$ from the MSD data (as in Fig.~\ref{fig2}), even so, such procedure would be equivalent to the aforementioned approach based on $\eta^{\prime}(\omega)$, this because the MSD and the compliance are related through Eq.~(\ref{compliance}) and, as shown in Fig.~\ref{fig3}(a), one have that $J(\tau) = \tau/\eta_s^{\,}$ for $\tau \ll \tau_0$ and $J(\tau) = \tau/\eta_0^{\,}$ for $\tau \gg \tau_f$.

	It is worth mentioning that, since $\eta_r^{\,}$ should depend on both the number density $n_f$ and the molecular weight $M_f$
of the filaments~\cite{rubinstein}, Eq.~(\ref{relative_visco}) tells us that, at least for dilute solutions, the number of beads $N$ of the EFM should also present a similar dependence on those quantities.~Here 
	we recall that the concentration of filaments $w_{f}$ (given in \% w/w) is related to the number density as
$w_{f} = n_{f} M_{f}/(n_{s} M_{s} + n_{f} M_{f})$,
where $n_s$ and $M_{s}$ are the number density and molecular weight of the
solvent molecules, respectively.
	Hence, by considering that the molecular weight of PAM chains is $M_{\text{PAM}}=18 \times 10^6\,$g/mol (see Ref.~\cite{tassieri2010jrheol}), the number density and the molecular weight of water molecules are, respectively, $n_{\text{water}}=3.34 \times 10^{22}\,$cm$^{-3}$ and $M_{\text{water}}=18\,$g/mol, and that the concentration used in the experiments~\cite{tassieri2010jrheol}
was $w_{\text{PAM}}^{\,}= 0.07$\% w/w, one finds that the number density of PAM chains is $n_{\text{PAM}}^{\,}\approx 2.33 \times 10^{13}\,$cm$^{-3}$.
	By assuming a relaxation modulus $G(\tau)$ similar to the one defined by Eq.~(\ref{relaxation_modulus_f}) one can evaluate
the low-frequency viscosity $\eta_0^{\,}$ of dilute solutions through Eq.~(\ref{eta0}), which yields
\begin{equation}
\eta_0^{\,} 
\approx
\frac{4}{3}
n_{f} \,k_B T \, \tau_f ~~.
\label{viscosity_nf}
\end{equation}
	Although a similar expression can be obtained specifically for flexible filaments~\cite{rubinstein,doisoftmatter}
by considering $G(\tau)$ defined by Eq.~(\ref{relaxation_modulus_f}) with~$\alpha=1/2$,
	it seems that approximated expressions, {\it i.e.}, with slightly different pre-factors, should be valid for complex fluids in general~\cite{rizzi2020jrheol}.~Indeed, 
by assuming that
$\eta_0^{\,} \approx 460\,$mPa.s, 
$\tau_f^{\,} \approx 3.38\,$s, and that the experiments with PAM~\cite{tassieri2010jrheol} were done at $T=25^{\text{o}}$C ({\it i.e.},~$k_BT=4.114\,$pN.nm),
Eq.~(\ref{viscosity_nf}) yields $n_f \approx 2.5 \times 10^{13}\,$cm$^{-3}$, 
which is in good agreement with the value of $n_{\text{PAM}}^{\,}$
estimated from the molecular weights.

	Now, with the values of $\eta_0^{\,}$, $\eta_s^{\,}$, 
and $\tau_f$ estimated from the experiments one could determine, at least in principle, the value of the elastic constant 
$\kappa$ through Eq.~(\ref{tauf_Rouse}) by considering Eqs.~(\ref{longtime_diff_coeff}) and~(\ref{relative_visco}), that is,
\begin{equation}
\kappa 
= \frac{6 \eta_0^{2}}{\pi \eta_s \tau_f} \, a  ~~.
\label{kappa_of_a}
\end{equation}
	The obtained expression shows that the elastic constant display a dependence on the radius $a$ of the probe particle, which is the only arbitrary (free) parameter of the EFM.~However, 
since $\zeta$ (or $D_0$) will also depend on $a$ as in Eq.~(\ref{selfdifcoef}), one can verify that, for both numerical methods
(i.e, overdamped dynamics and relaxation simulations), the resulting MSD will be proportional to $a^{-1}$ so that the 
compliance $J(\tau)$ evaluated via Eq.~(\ref{compliance}) will not depend on the value of $a$.~Interestingly, 
	such ``renormalizability'' can be seen as a suitable feature of our methodology since it occurs just as expected from any meaningful microrheological approach.~It 
	can be instructive to consider Eqs.~(\ref{longtime_diff_coeff}) and~(\ref{relative_visco})
in order to replace $a$ in Eq.~(\ref{compliance}) 
so that the compliance can be rewritten as
\begin{equation}
J(\tau) = \frac{1}{2 d \, \, D_f \, \eta_0^{\,}} \langle \Delta r^2 (\tau) \rangle  ~~.
\label{Jtau_nf}
\end{equation}
	Even so, Eq.~(\ref{longtime_diff_coeff}) indicates that one still have an implicity dependence of $D_f$ on $a$, which will be cancelled out by the implicity dependence of the MSD on $a$ as well.
	In fact, the restriction on the values of the EFM parameters is imposed by experimental data mainly through Eqs.~(\ref{tauf_Rouse})~and~(\ref{relative_visco}), which only require that the product of the elastic constant and the diffusion coefficient have a specific value, {\it i.e.}, $\kappa D_0 = k_BT \eta_0^{2}/\pi^2 \tau_f\eta_s^{2}$.
	For example, in order to obtain the compliance $J(\tau)$ 
that is displayed in Fig.~\ref{fig3}(a),
we choose $\kappa D_0 =12.342\,$pN.nm/s, which is consistent 
with the values of $T$, $\eta_0^{\,}$, $\eta_s^{\,}$, and 
$\tau_f$ that were estimated from experiments.
	The ambiguity in the definitions of $D_0$ (or $D_f$), Eq.~(\ref{longtime_diff_coeff}), and $\kappa$, Eq.~(\ref{kappa_of_a}), can be eliminated only when one set an arbitrary value to the radius of the probe particles, {\it e.g.},~$a=2.8\,\mu$m as used in Ref.~\cite{tassieri2010jrheol}, which yields 
 $D_0 = 1.69 \,$nm$^2$/ms and
$\kappa = 7.28\times 10^{-3}\,$pN/nm.

	In addition, we note that there is some freedom when choosing the physical units of the time increment $\Delta t$ used in the simulations (see Eqs.~(\ref{langevin_integrated_brownian}) and~(\ref{langevin_integrated})), and usually it is convenient to set it with the same time units that is used to define $D_0$.
	However, it is, in fact, more important that its value is small enough 
so that it ensures not only numerical stability but also that a free-like diffusion regime with the MSD given by $\langle \Delta r^2 (\tau) \rangle \approx 2d D_0 \tau$ is observed. For instance,
	in order to obtain the numerical data displayed in Fig.~\ref{fig3}(a) we consider 
$\Delta t=0.1\,$ms, which is much shorter than $\tau_0 \approx 33.8\,$ms
(and it is not displayed in the figure).

	Accordingly, as shown in Fig.~\ref{fig3}(a), the EFM defined with parameters that were 
determined from a few limiting values allowed us to show how one can apply the relaxation simulations to obtain the non-markovian dynamics that lead to the whole compliance function $J(\tau)$ that agreed with the experimental curve.

	Next, we consider Eq.~(\ref{complexG_J}) to obtain
the complex shear modulus $G^{*}(\omega)$ from $J(\tau)$, where the Fourier
 transform of the compliance $\hat{J}(\omega)$ is evaluated numerically by
the method proposed in Ref.~\cite{evans2009pre} 
(see Ref.~\cite{tassieri2012newj} for further details).~In addition, 
	we evaluate the complex viscosity from $G^*(\omega)$ as
\begin{equation}
\eta^*(\omega)=\dfrac{G^*(\omega)}{i\omega}~~.
\label{complex_viscosity}
\end{equation}
	Figures~\ref{fig3}(b) and~\ref{fig3}(c) indicates that 
both storage $G^{\prime}(\omega)$ and loss $G^{\prime \prime}(\omega)$ modulus, as well as the viscosity $\eta^{\prime}(\omega)=G^{\prime \prime}(\omega)/\omega$, present a surprisingly good agreement to the experimental data obtained from PAM chains~\cite{tassieri2010jrheol}.
	In particular, Fig.~\ref{fig3}(b) shows that, at
low frequencies, 
$\omega \ll \tau_f^{-1} \approx 0.3\,$rad/s,
$G^{\prime}(\omega) \propto \omega^2$
and  $G^{\prime \prime}(\omega) \propto \omega$, which
means that the viscosity
$\eta^{\prime}(\omega)=G^{\prime \prime}(\omega)/\omega$ 
goes to a constant value,
$\eta_{0}^{\,} \approx 460\,$mPa.s.~As 
	shown in Fig.~\ref{fig3}(c), $\eta^{\prime}(\omega)$ tends to a value $\eta_{0}^{\,}/N$ at high frequencies,
thus, as mentioned earlier, one can consider those values to estimate the effective number of beads $N$
of the EFM through Eq.~(\ref{relative_visco}).


	Here, it is important to emphasize that  
Eqs.~(\ref{longtime_diff_coeff})-(\ref{Jtau_nf}), 
which were obtained for flexible chains, are also valid for an EFM defined with $\kappa_b \neq 0$.~Hence, 
as we discuss below, the same approaches presented in this section can be applied in the study of
dilute solutions of semiflexible filaments.



\section{Results}
\label{semiflexible}

	In the following we present numerical results obtained for solutions of semiflexible filaments.~In
	particular, we analyse the effect of bending energies on the 
dynamics of EFMs characterized by both short and long filaments, demonstrating the effectiveness of our relaxation
simulations described in Sec.~\ref{relax_sim}.

\begin{figure*}[!ht]
	\centering
	\includegraphics[scale=0.6]{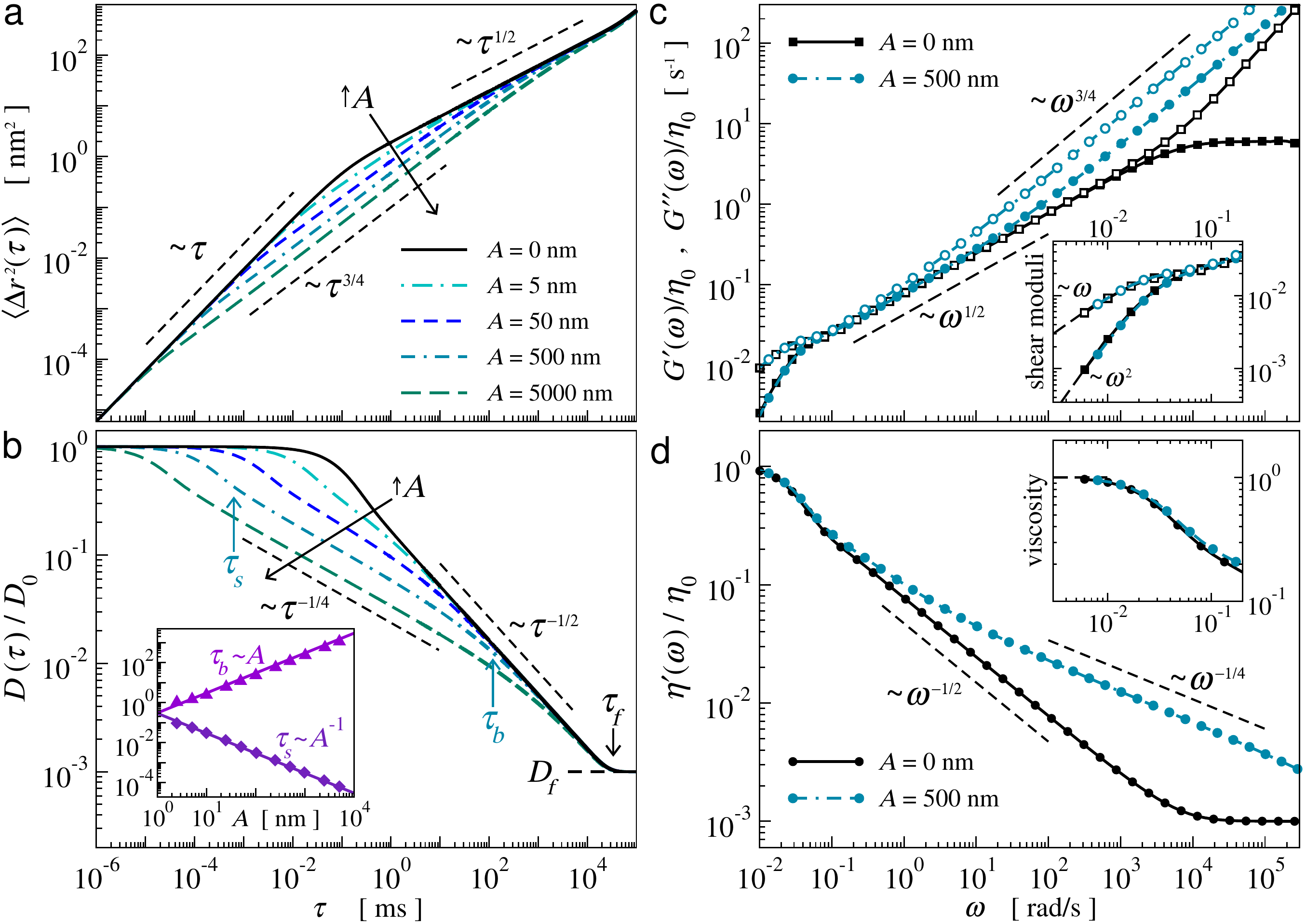}
	\caption{
	Effect of bending energies on the dynamics of semiflexible EFMs 
and the viscoelastic properties of the 
corresponding solutions.
	Here we labelled the data in terms of $A=E^{\prime}/b$, which should be proportional to both the bending constant $\kappa_b = E' k_B T/b^4$ and the persistent length $L_p$ of the EFM.
	(a)~Mean-squared displacement of the $m$-th bead $\langle \Delta r^2(\tau)\rangle$, Eq.~(\ref{deltaR2relax}); (b)~time-dependent diffusion coefficient $D(\tau)$, Eq.~(\ref{relaxDtau}) (Inset: characteristic times $\tau_s$ and $\tau_b$ as a function of $A$); (c) 
reduced 
 storage modulus  $G^{\prime}(\omega)/\eta_0^{\,}$ (filled symbols), and loss modulus $G^{\prime \prime}(\omega)/\eta_0^{\,}$ (open symbols), obtained from the compliance $J(\tau)$, 
 Eq.~(\ref{Jtau_nf}), 
through Eq.~(\ref{complexG_J}); and (d) 
reduced 
 viscosity, $\eta^{\prime}(\omega)/\eta_0^{\,}$, Eq.~(\ref{complex_viscosity}).
	Inset panels of (c) and (d) are just zoomed in regions to show the low frequency
regimes.~In order to clearly demonstrate both the effects of bending and the efficiency of our approach, the relaxation simulations were implemented for EFMs defined
with $N=1000$  
and different values of the bending constant $\kappa_b$.~Results were obtained with a constant external force $F_0=1\,$pN, 
$D_0=1\,$nm$^2$/ms,
$T =300\,$K ($k_BT=4.142\,$pN.nm),  
$\Delta t=10^{-6}\,$ms,
and with a fixed value for the elastic constant $\kappa = 12.426\,$pN/nm, 
so that $\tau_f \approx 3.4 \times 10^4\,$ms for all $\kappa_b$ (or $A$). 
	Short-dashed (black) lines indicate power-law behaviours observed for the dynamical and rheological quantities.}
	\label{fig4}
\end{figure*}

	Figure~\ref{fig4} include results obtained for dilute solutions of filaments described by semiflexible EFMs
 composed by $N=1000$ beads defined with different values for the bending constant $\kappa_b$ but with a fixed elastic constant $\kappa$.
	The results presented in Fig.~\ref{fig4} were obtained for long chains with parameters that were arbitrarly chosen
in order not only to demonstrate the usefulness of our relaxation simulations, but also to show the effects of bending energies on time scales that are clearly distinguishable on the dynamical quantities, {\it i.e.}, $\langle \Delta r^2(\tau) \rangle$ and $D(\tau)$, and on the viscoelastic functions, {\it i.e.}, $G^{\prime}(\omega)$, $G^{\prime \prime}(\omega)$, and $\eta^{\prime}(\omega)$.
	Importantly, we labelled the results in terms of $A \equiv E^{\prime}/b$, which is a quantity that is directly related to the bending constant, as $\kappa_b=E/b^4$ and $E^{\prime}=E/k_B T$ (see Sec.~\ref{filamentmodel}),~ and also because the persistent length $L_p$ of the EFM should be proportional to $A$.
	In practice, higher values of $\kappa_b$ correspond to greater values of $A$, and the corresponding effective media can be interpreted as EFMs having longer persistent lengths $L_p$.
	Nevertheless, one should note that the EFM is used to describe the
effective bending energy that result from the coupling between a probe particle and 
the semiflexible filaments in solution, thus $L_p$ should represent an effective quantity
rather than the persistent length of a single semiflexible filament in solution.~Here, 
in order to have numerical values attributed to $A$ (hence to $\kappa_b$), we simply choose $k_BT=4.142\,$pN.nm ($T=300\,$K), $D_0=1\,$nm$^2$/ms, and $b=1\,$nm.~These 
	choices set values to the elastic constant, $\kappa=3k_BT/b^2=12.426\,$pN/nm, 
and to the longest relaxation time, Eq.~(\ref{tauf_Rouse}), $\tau_f \approx 3.4 \times 10^4\,$ms, but
not to the viscosity $\eta_0^{\,}$, so we present just the reduced viscoelastic functions in Figs.~\ref{fig4}(c) and~\ref{fig4}(d).

	Figures~\ref{fig4}(a) and~\ref{fig4}(b) show that the bending energy lead
to significant changes in the dynamics of the beads in the EFM.
	In particular, the short-time diffusion dynamics observed for the flexible EFM ($A=0\,$nm)
is altered to an extended subdiffusive regime where $\langle \Delta r^2(\tau) \rangle \propto \tau^{\alpha}$,
with $\alpha$ approaching $3/4$ as the value of $A$ increases.
	This behaviour is confirmed in Fig.~\ref{fig4}(b) by the time-dependent diffusion coefficient $D(\tau)$,
from where one can verify that the shortest relaxation time 
decreases as $A$ (and $\kappa_b$) increases, while 
changes in the bending constant $\kappa_b$ seems to not alter the longest relaxation time $\tau_f$ (at least for $A < 5000\,$nm).
	Figure~\ref{fig4}(b) indicate that higher values of $A$ lead to a wider range of subdiffusive anomalous behaviour where $D(\tau) \propto \tau^{-1/4}$.
	By considering a local power-law approximation for the time-dependent diffusion coefficient, {\it i.e.},~$D(\tau) \propto \tau^{\nu}$, we computed the numerical derivatives of $\nu$ and, from its inflexion points, we determine two characteristic time scales, $\tau_s$ and $\tau_b$, which comprise the range of subdiffusive behaviour that is directly related to the introduction of the bending energy, as illustrated for $A=500\,$nm in Fig.~\ref{fig4}(b).
	Interestingly, our results indicate that both characteristic times depend on $A$ in a simple way and, as shown in the inset of Fig.~\ref{fig4}(b), 
$\tau_s \approx 0.3 \, A^{-1}$ and $\tau_b \approx 0.3 \, A$.
	In addition, we observe that, at least for that range of $A$, the long time diffusion coefficient $D_f=D_0/N$ and the longest relaxation time $\tau_f$ remained unaltered, thus they can be conveniently evaluated from Rouse estimates through Eqs.~(\ref{longtime_diff_coeff}) and~(\ref{tauf_Rouse}), respectively.

\begin{figure}[!t]
	\centering
	\includegraphics[scale=0.57]{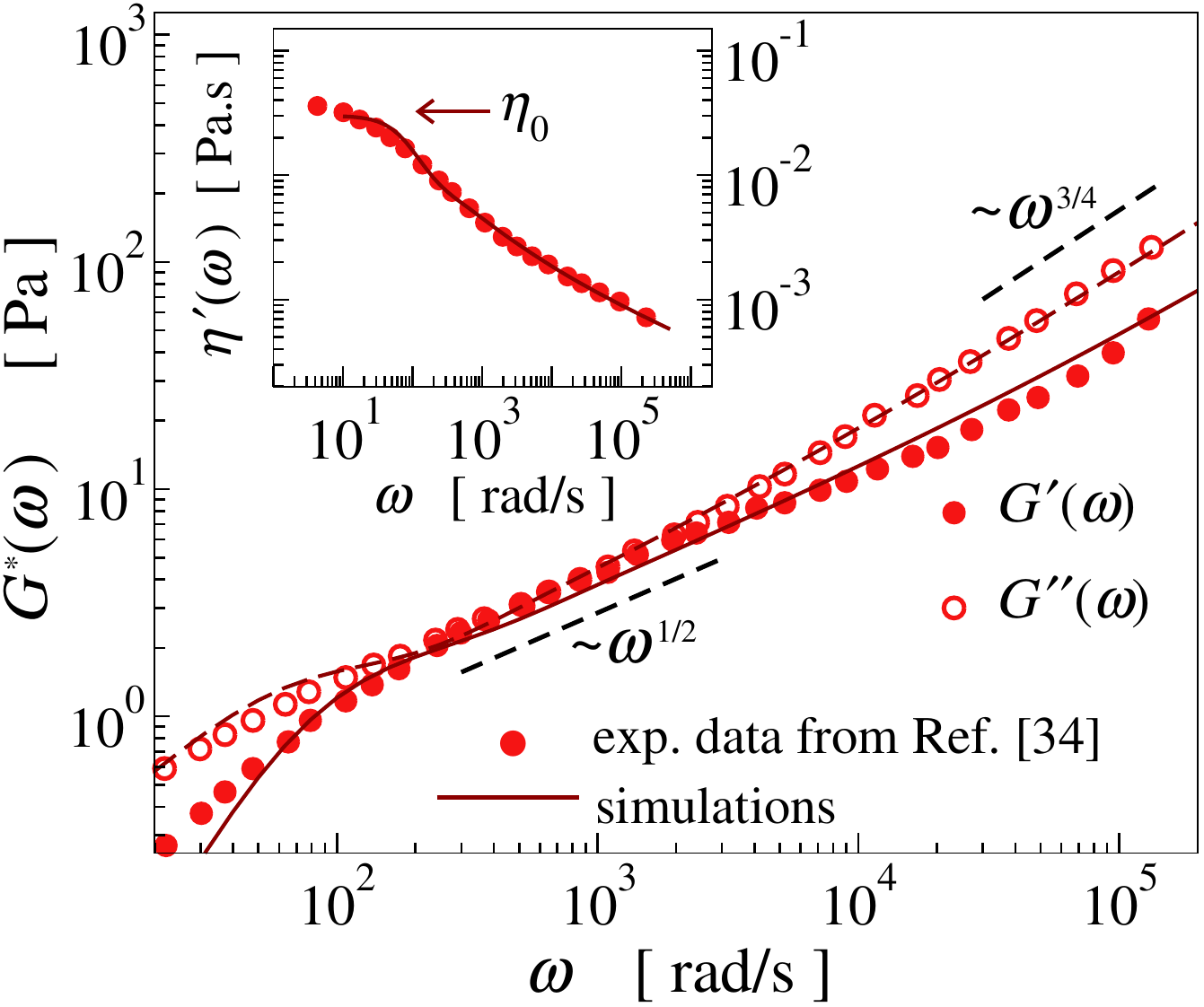}
	\caption{
	Comparison between the shear moduli, $G^{\prime}(\omega)$ and $G^{\prime \prime}(\omega)$, obtained from relaxation simulations (continuous lines), as described in Sec.~\ref{visco_prop}, and the experimental data (circles) obtained for a
$1\,$mg/mL $5.8$\,kilobase DNA solution~\cite{krajina2017acs}.
	The inset panel displays the frequency-dependent viscosity evaluated from the loss modulus, $\eta^{\prime}(\omega) = G^{\prime \prime}(\omega)/\omega$.
	Relaxation simulations were performed with 
$F_0=1\,$pN, 
$N=1000$, 
$D_0=14.45\,$nm$^2$/$\mu$s, 
$\kappa=2.2\,$pN/nm ($b \approx 2.36\,$nm),   
 $\kappa_b=367.28\,$pN/nm ($A \approx 1179\,$nm), 
and $\Delta t=10^{-6}\,\mu$s, 
in order to be consistent with the given experimental conditions, {\it i.e.},
$\eta_0^{\,} \approx 30\,$mPa.s,  
$T\approx 23^{\text{o}}$C ($k_B T=4.086\,$pN.nm), 
and
$\tau_f \approx 0.013\,$s.~The
	values of $D_0$ and $\kappa$ were determined respectively through Eqs.~(\ref{longtime_diff_coeff}) and~(\ref{kappa_of_a})
by setting $a=0.5\,\mu$m, which is the radius of the probe particles specified in Ref.~\cite{krajina2017acs}.~By
	 considering Eq.~(\ref{viscosity_nf}), the estimated number density of DNA molecules in solution is equal to $n_{f}^{\,} \approx 42 \times 10^{13}\,$cm$^{-3}$, which yields a molecular weight of $1.42 \times 10^6\,$g/mol, in a reasonable agreement with the size and the concentration of DNA used in the experiments.
}
	\label{fig5}
\end{figure}

	As shown in Figs.~\ref{fig4}(c) and~\ref{fig4}(d), the changes in the short time dynamics of
the beads in the semiflexible EFM clearly modify the high frequency viscoelastic response of the solution.
	In contrast to the dynamics of the beads in the flexible EFM, which display a characteristic exponent $\alpha = 1/2$ at high frequencies~\cite{doiedwards} (see, {\it e.g.}, Fig.~\ref{fig2}), greater values of $A$ lead to
a subdiffusive anomalous behaviour so that the
reduced moduli are given by $G^{\prime}(\omega)/\eta_{0}^{\,} \propto \omega^{\alpha}$ and $G^{\prime \prime}(\omega)/\eta_{0}^{\,} \propto \omega^{\alpha}$, and the relative viscosity is given by $\eta^{\prime}(\omega)/\eta_{0}^{\,} \propto \omega^{\alpha-1}$, with a characteristic exponent $\alpha \approx 3/4$, in agreement to what have been suggested by previous theoretical and 
computational studies presented in the literature~\cite{rubinstein,binder2016jcp,shankar2002jrheol,pasquali2001pre,dimitrakopoulos2001pre,spakowitz2014pre}.
	Also, as expected from the dynamics, one can verify from the inset of Figs.~\ref{fig4}(c) and~\ref{fig4}(d) that the low frequency regimes of the reduced moduli and the reduced viscosity are not altered due to the introduction of the bending energy.

	A careful look at Fig.~\ref{fig4}(c) indicates that, in addition to
the power-law behaviour observed for the shear moduli with a somewhat characteristic 
exponent $\alpha=3/4$, the rheology of solutions of semiflexible filaments might display 
intermediate values of $\alpha$, also including a transition regime from the 
flexible behaviour with exponent $\alpha=1/2$.
	In order to illustrate that ideia we include comparisons between the numerical results
obtained from our relaxation simulations and experimental data extracted from microrheology experiments.

	For instance, Fig.~\ref{fig5} shows the viscoelastic response obtained for a dilute solution 
of DNA~\cite{krajina2017acs} which is very similar to the behaviour
observed for intermediate values of $A$ and long filaments showed in Fig.~\ref{fig4}(c), that is, a transition from a subdiffusive regime with $\alpha\approx 1/2$ at intermediate frequencies to a regime where 
$G^{\prime}(\omega) \sim G^{\prime \prime}(\omega) \sim \omega^{3/4}$, at high frequencies.
	Unfortunately, as shown in the inset of Fig.~\ref{fig5}, the frequency-dependent 
viscosity $\eta^{\prime}(\omega)$ obtained from the experiments do not include data at 
sufficiently high frequencies which would have allowed us to determine the exact number of 
beads $N$ of the EFM via Eq.~(\ref{relative_visco}).
	Even so, by arbitrarly choosing $N=1000$ as well as by considering 
$\kappa D_0=31.846\,$pN.nm/$\mu$s 
in order to match
the experimentally estimated viscosity, $\eta_0^{\,} \approx 30\,$mPa.s, 
temperature, $T=23^{\text{o}}$C, and
longest relaxation time, $\tau_f \approx 0.013\,$s, 
we were able to attain a quantitative agreement between the numerical results and the experimental data.
	Here, it is worth noting that the viscosity $\eta^{\prime}(\omega)$ displayed in the inset of Fig.~\ref{fig5} 
spams several orders of magnitude, so that, according to Eq.~(\ref{relative_visco}), only large values of $N$ would be suitable to describe the experimental data.

\begin{figure}[!b]
	\centering
	\includegraphics[scale=0.36]{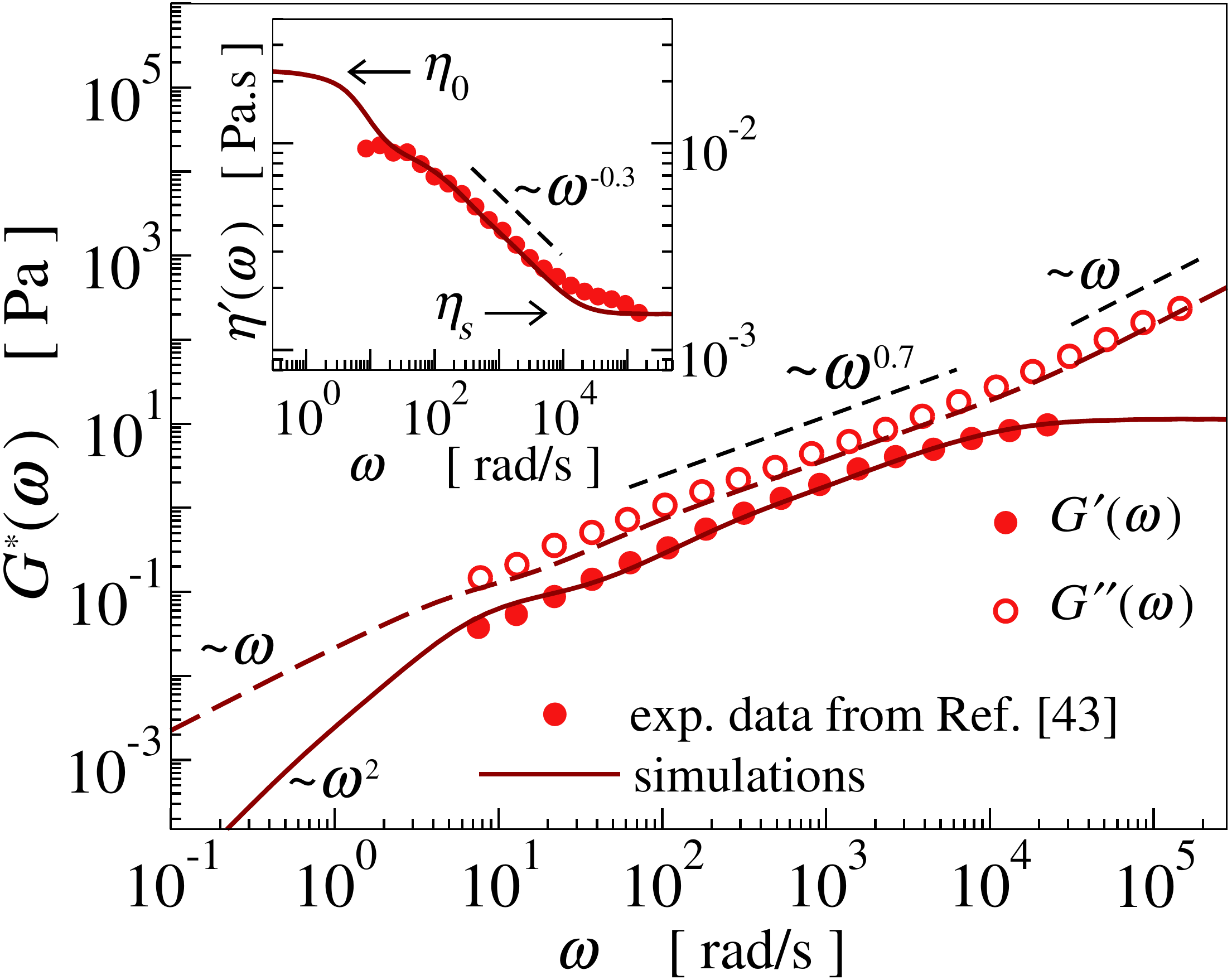}
	\caption{
	Comparison between the viscoelastic properties obtained from relaxation simulations (lines), as 
described in Sec.~\ref{visco_prop}, and the experimental data (circles) extracted from Ref.~\cite{shayegan2013plosone}, which corresponds to solutions of collagen macromolecules at $2\,$mg/mL.
	The main panel shows the storage modulus (filled circles), $G^{\prime}(\omega)$, and the loss modulus (open circles), $G^{\prime \prime}(\omega)$, while the viscosity $\eta^{\prime}(\omega)=G^{\prime \prime}(\omega)/\omega$ is displayed in the inset panel.~Relaxation 
	simulations were performed with $F_0=1\,$pN,
$N=15$, 
$D_0=68.67\,$nm$^{2}$/ms, 
 $\kappa = 1.317 \times 10^{-3}\,$pN/nm ($b\approx 96\,$nm),  
$\kappa_b = 74.1 \times 10^{-3}\,$pN/nm ($A \approx 1.62 \times 10^4\,$nm),
and $\Delta t=10^{-6}\,$ms, 
in order to be consistent with the given experimental conditions, {\it i.e.},
$T\approx 21^{\text{o}}$C ($k_B T=4.059\,$pN.nm), 
$\eta_s^{\,} \approx 1.49\,$mPa.s, 
$\eta_0^{\,} = N \eta_s^{\,} \approx 22.4\,$mPa.s,
and 
$\tau_f \approx 1.023\,$s.
	The specific values of $\kappa$ and $D_0$ were determined 
by setting the radius of the probe particles equal to $a=2.1\,\mu$m as in Ref.~\cite{shayegan2013plosone}.
}
	\label{fig6}
\end{figure}

	Next, we include in Fig.~\ref{fig6} a comparison between the numerical results obtained from our relaxation simulations and the experimental data on solutions of collagen at $2\,$mg/mL extracted from Ref.~\cite{shayegan2013plosone}.
	Unfortunately, the experimental data of Ref.~\cite{shayegan2013plosone} do not include the low frequency
regime of the viscoelastic functions, even so we use the frequency-dependent viscosity $\eta^{\prime}(\omega)$ to estimate
the number of beads of the EFM as $N=15$, which gives $\eta_0^{\,} = N \eta_s^{\,} \approx 22.4\,$mPa.s, and we also
consider $\tau_f \approx 1.023\,$s, so that Eq.~(\ref{tauf_Rouse}) yields 
$\kappa D_0\approx 90.45\,$pN.nm/s.
	Although the effective number of beads is small as in the case of PAM, it is worth mentioning that, as in the case of DNA, several relaxation simulations were required to determine a suitable value for the bending constant $\kappa_b$ in order to obtain meaningful viscoelastic response functions over the full range of frequencies.~The 
	results presented in Fig.~\ref{fig6} indicate that
the exponent $\alpha$ observed for the power-law behaviour of the shear moduli 
present a value between $1/2$ and $3/4$ at an intermediate frequency range.
	The value of $\alpha \approx 0.7$ is corroborated by the behaviour of
the frequency-dependent viscosity $\eta^{\prime}(\omega)\propto \omega^{\alpha-1}$ which is displayed in the inset panel.~As 
	suggested in Ref.~\cite{shayegan2013plosone}, such intermediate behaviour between flexible and semiflexible could be explained due to ratio between the short contour length ($L\approx300\,$nm) and the persistent length ($L_p\approx 15-160\,$nm) of the collagen molecules, which would put the viscoelastic response of the corresponding solution in a crossover region.
	However, the number density of filaments in solution estimated through Eq.~(\ref{viscosity_nf}) is $n_{f}^{\,} \approx 0.03 \times 10^{13}\,$cm$^{-3}$, which is a very low value for the
	nominal concentration of $2\,$mg/mL.~This 
value of $n_f$ leads to a ``molecular'' weight of $2.2 \times 10^9\,$g/mol, indicating that the filamentous structures in solution might be, in fact, self-assembled fibers that are much larger than the $300\,$kDa collagen molecules assumed in Ref.~\cite{shayegan2013plosone}.
	Interestingly, the presence of large supramolecular structures composed by thousands of macromolecules might explain why the EFM required only a rather small effective number of beads 
in order to describe the viscoelasticity of dilute solutions of semiflexible collagen molecules.
	
	It is worth noting that, in general, the effective number of beads $N$ of the EFM is constrained by the experimentally available information through Eq.~(\ref{relative_visco}), but one can choose any arbitrary value for the radius $a$ of the probe particles (see Sec.~\ref{visco_prop}).~Although the viscoelastic functions $J(\tau)$, $G^{\prime}(\omega)$, and $G^{\prime \prime}(\omega)$ obtained from our approach do not depend on the value of $a$, 
it is advisable to restrict it to the micron-sized range as in the most of microrheological experiments~\cite{rizzi2018eac} in order to avoid unphysical values for the effective coupling constants $\kappa$, Eq.~(\ref{kappa_of_a}), and $\kappa_b$.
	In addition, we emphasize that the application of the EFM and the FDT should be limited to cases where microrheological approaches are valid, in particular, when the so-called Stokes and Einstein components
are valid (see Ref.~\cite{squires2010annrev}).


\section{Concluding remarks}

	In this work we present efficient relaxation simulations as an alternative numerical method
that allow one to evaluate the viscoelastic response of both flexible and semiflexible filament solutions.
	In particular, our study indicates that the evaluation of the MSD and the time-dependent diffusion coefficient of the probe particles described by the EFM is orders of magnitude quicker for the relaxation simulations in comparison to the stochastic approach.
	In addition, by considering a modelling approach that is based on microrheology,
we have established useful relations that allowed us to obtain the shear moduli and the viscosity of unentangled filament solutions without having to resort to shearing protocols that are used for solutions with cross-linked and tighly entangled filaments.

	Despite of the fact that we have used a simple modelling approach, the quantitative agreement 
between our numerical results and the experimental data presented in Figs.~\ref{fig5} and~\ref{fig6} demonstrate
the effectiveness of the EFM in obtaining meaningful frequency-dependent viscoelastic response functions for dilute solutions of semiflexible filaments.
	We believe that it could encourage the use of the relaxation methodology combined with more detailed 
approaches.
	In particular, it might be interesting to extend the EFM to incorporate excluded volume effects in order to describe other physical scenarios, {\it e.g.}, semi-dilute solutions~\cite{sarmiento2012EPJE,chen2011jrheol}.
	Also, since the relaxation simulations have been previously adapted to obtain the dynamics of random flexible polymers networks through an averaging procedure~\cite{teixeira1999epl}, one might also extend it to more complex solutions which display locally distributed viscoelastic properties~\cite{rosa2018jcp,azevedo2020jphysconfser,rizzi2020jrheol}.
	Even so, one must realize that although the approach described in Sec.~\ref{visco_prop} is useful to 
provide estimates for the parameters $N$, $\kappa$, and $D_0$, the EFM is not a microscopic, {\it i.e.}, molecular-based, model, hence it lacks the predictive power that one may desire in many real-life applications.~For solutions of semiflexible polymers, in particular,
	there might be still the need to perform a 
considerable large number of simulations in order to test EFMs defined with different values of the bending 
constant $\kappa_b$ until one obtain appropriate viscoelastic functions, {\it i.e.}, $G^{\prime}(\omega)$, 
$G^{\prime \prime}(\omega)$, and  $\eta^{\prime}(\omega)$, in a wide range of frequencies.~Nevertheless, this also
	emphasizes the importance of the efficiency of our relaxation approach presented in
Sec.~\ref{relax_sim} when 
compared to the stochastic simulations described in Sec.~\ref{stochastic_sim_sec}.

	Finally, it is worth mentioning that, since the theoretical basis of the relaxation simulations is the fluctuation-dissipation theorem, the determination of the mean-squared displacement and the time-dependent diffusion coefficient of the beads through Eqs.~(\ref{deltaR2relax}) and~(\ref{relaxDtau}), respectively, does not need to be based neither on stochastic simulations or on the EFM to be accomplished.~Hence,
	 one might try to explore those equations together with Eqs.~(\ref{compliance}) and~(\ref{complexG_J}), 
just as it has been done in microrheology experiments~\cite{tassieri2010jrheol}, and associate them 
with molecular-based simulations in order to provide the dynamics of a probe particle which takes into account its interactions with all solvent and polymeric molecules in solution.
	Alternatively, one may consider systematic coarsening procedures based also on more fundamental ({\it i.e.},  molecular-based) simulations~\cite{gartner2019macromol} which could be used, for instance, to
develop improved EFMs by establishing effective potentials for which the dynamics of the probe particle is affected by specific microscopic conditions of the polymers in solution.




	L. G. Rizzi acknowledges the financial support of the Brazilian 
agencies CNPq (Grants N\textsuperscript{o} 306302/2018-7 and N\textsuperscript{o} 426570/2018-9) and FAPEMIG (Process APQ-02783-18), and L. K. R. Duarte thanks the scholarship from the Brazilian agency CAPES.







\vspace{-0.3cm}

\section*{References}

\vspace{-0.65cm}




%

\end{document}